\documentstyle[12pt,fleqn,epsf]{article}
\def\ie{\hbox{\it i.e. }}        
        
\def\etal{\hbox{\it et al.}}

\def\ket#1{\left| #1\right\rangle}

\def\abs#1{\left| #1\right|}

\def\etap{\eta^\prime}
\def\beq{\begin{equation}}
\def\eeq{\end{equation}}
\def\bea{\begin{eqnarray}}
\def\eea{\end{eqnarray}}

\def\lra{\longrightarrow}

\def\com#1#2 {\left[#1, #2\right]}
\def\np#1#2#3{    {\it Nucl. Phys. }{\bf #1} #2 (19#3)}
\def\pl#1#2#3{    {\it Phys. Lett. }{\bf #1} #2 (19#3)}
\def\pr#1#2#3{    {\it Phys. Rev. }{\bf #1} #2 (19#3)}

\def\prl#1#2#3{    {\it Phys. Rev. Lett. }{\bf #1} #2 (19#3)}

\def\zp#1#2#3{    {\it Zeit. f\"ur Physik }{\bf #1} #2 (19#3)}

\def\ijm#1#2#3{   {\it Int. J. Mod. Phys. }{\bf #1} #2 (19#3)}

\def\nc#1#2#3{     {\it Nuovo Cim. }{\bf #1} #2 (19#3)}

\def\mxfigura#1#2#3#4{
  \begin{figure}[htp]
    \begin{center}
      \epsfxsize=#1
      \leavevmode
      \epsffile{#2}
     \end{center}
    \caption{#3}
    \label{#4}
  \end{figure} }
\def\prepnuma{FTUV/94-55}
\def\prepnumb{IFIC/94-50}

\def\titol{$\eta -\eta^\prime$
Photoproduction and \\the Axial Isoscalar Neutral Current Coupling}
\def\autora { J.  Bernab{\'e}u, G.A.  Gonz{\'a}lez-Sprinberg and J.
Vidal}
\def\adressaa{Departament de F{\'{\i }}sica Te{\`o}rica,
Universitat de Val{\`e}ncia, \\ and IFIC, Centre Mixt Univ.
Valencia-CSIC\\ E-46100 Burjassot (Val\`encia), Spain}

\def\resum{We show that coherent $\eta$ and $\etap$ photoproduction by
 means of the Primakoff Effect on the proton depends on the strange
 component of the neutral axial current coupling.  We construct
 polarization asymmetries that are sensitive to this coupling through
 the $\gamma - Z$ interference.  The $\eta^\prime$ is not a
 Goldstone boson of a spontaneously broken chiral symmetry,
 but a phenomenological analysis of the $\eta$ and
 $\eta^\prime$ production through chiral perturbation theory allows to
 calculate the observables of interest.  The polarized proton or
 polarized photon
 asymmetries are predicted to be close to $10^{-4}$ for $-q^2 \sim
 0.1-0.5\;\mbox{\rm GeV}^2$.}

\def\firstpage{\begin{titlepage}
\baselineskip 0.50cm \null
\vspace*{-1cm}
\hfill \prepnuma\\
\null \hfill \prepnumb\\
\baselineskip 0.75cm
\vskip 3.cm
\begin{center}
{\Large \bf \titol}
\vskip 2.cm
\baselineskip 0.6cm
{\bf \autora}
\vskip 1.cm
{\it \adressaa}
\vskip 1.5cm
{\bf ABSTRACT}
\end{center}
 \baselineskip 0.60cm
 \begin{quotation}
  \resum
  \end{quotation}

\end{titlepage}
\baselineskip 0.75cm}
\begin{document}
\firstpage
\setcounter{page}{2}

\def\tendeix#1#2#3{ \stackrel{#2\rightarrow #3}{\longrightarrow} #1}
      \section{Introduction} The flavour content of the nucleon has
      been experimentally tested and confronted with effective
      theories on the nucleon structure all over the past 20-30 years.
      So far, a rather consistent picture of the up and down content
      of the nucleon has emerged from both low and high energy
      experiments, whereas predictions from the quark-type models,
      Skyrme-inspired models and the quark-parton model have been
      extensively tested.  However, a deep understanding of the
      nucleon structure starting from the believed strong interaction
      theory, namely QCD, is still lacking.  The relation of the
      quark-type models to low energy QCD is obscure, and the
      Skyrme-type effective theories have little evolved since the
      fundamental papers of their revival in the '80.

An open window to new phenomena related to the flavour content of the
nucleon started since the EMC and SLAC deep inelastic scattering
experiments \cite{emcslac}.  From the results of these, the common
wisdom of the (naive) spin structure of the proton has given rise to a
lively debate.  Data from experiments, when combined with the analysis
of semileptonic baryon decays and the quark-parton model, give the
polarized quark moments $\Delta q \; (q=u,d,s)$ in the proton.  The
flavour singlet part of these first moments is found to be anomalously
small, leading to the so called `spin crisis'.  Moreover, the strange
polarization quark moment is predicted to be of the same order of
magnitude of the up and down ones.

Another probe of the proton flavour content is provided by the weak
neutral axial current.  This current certainly receives contributions
from the up and down quarks, and the previously mentioned experiments
that lead to the `spin crisis' raise the question on the strange
content of the proton and neutron.  So the following question imposes:
is it possible to have a strange flavour contribution comparable with
the up and down ones?  In other words, is it possible to have an
isoscalar neutral axial current coupling comparable with the isovector
one?  This is the question to which we address to in this paper.

Our aim is to construct observables sensitive to the axial isoscalar
coupling of the proton.  This can be searched for \cite{bern} in
elastic neutrino-proton scattering or electroweak nuclear processes.
In Ref.\cite{nos} we have demonstrated that the polarized Primakoff
Effect is adequate to achieve this purpose.  In this process there
exists a neutral weak current contribution through the $\gamma
-Z-\pi^0$ vertex.  This contribution is suppressed by the factor $G_F
Q^2/\alpha$, relative to the pure electromagnetic one, and one has to
look for parity-violating asymmetries in order to disentangle it.  The
P-odd observables are induced by the weak- electromagnetic
interference for polarized photon or polarized proton.  The parity
violating asymmetries for polarized photon or polarized proton
Primakoff Effect ($\pi^0$ photoproduction) filter the couplings so as
to leave the proton neutral {\it axial} coupling only.  However, in
that case, in addition to the suppression factor $G_F Q^2/\alpha$, the
anomaly cancellation condition in the standard theory forces the
vector coupling of the electron to come into the game, and another
suppression factor appears in the asymmetries.  One could have
na\"{\i}vely expected a higher value because only $u$ and $d$ quarks
vector-couplings ($v^u$ and $v^d$) to $Z$ are needed in the calculus.
But the anomaly cancellation imposes $N_c (Q_u v^u -Q_d v^d) = Q_e
v^e$, and one ends up with the small factor $v^e$ in the asymmetries.
In the same way, assuming exact flavour-$SU(3)$ symmetry, in the case of
coherent photoproduction of the other flavour-neutral pseudoscalar
 meson in the octet,
the $\eta$, the same suppression factor is present.  We would like to
think of the possibility to avoid this suppression by considering the
coherent photoproduction of $\eta^\prime$.  What about the
$\eta^\prime$?  Mostly identified with the singlet component of the
flavour $U(3)$ meson nonet, it is not a Goldstone boson for any
of the symmetries of QCD, and in the zero quark mass limit
the axial $U(1)$ anomaly prevents
its mass to vanish.  So a similar analysis to the one we
developed in Ref.\cite{nos} is not possible for the $\eta^\prime$
meson.

In this paper we extend the above mentioned results so as to include
the $\eta$ and $\eta^\prime$ photoproduction as a probe of the
isoscalar axial coupling.  The absence of the anomaly cancellation
suppression factor already mentioned allows to obtain two orders of
magnitude enhanced asymmetries, thus reducing the need of
statistics by four
orders of magnitude.  The paper is organized as follows:  in section
(2) the theoretical basis of our computation is established; in
section (3) the announced observables are constructed, and in section
(4) we present the numerical estimates and discussion of the results.

\section{$\eta$, $\eta^\prime$ and the Primakoff Effect}

The spin-dependent structure function $g_{1} (x, Q^{2})$ of the
proton, as determined by the EMC-experiment together with previous
SLAC data for electron scattering, and the analysis of semileptonic
baryon decays give the polarized quark moments $\Delta q \;(q=u,d,s)$
in the proton:
\begin{eqnarray}
\Delta u & = & 0.78 \pm 0.06
\nonumber \\ \Delta d & = & -0.47 \pm 0.06 \\ \Delta s & = & 0.19 \pm
0.06 \nonumber
\end{eqnarray}
In particular, the flavour singlet part
of this first moment is found to be anomalously small,
leading to the
so called `spin crisis'.  Another by-product of this analysis is that
an unexpected large strange quark moment is obtained.

Another probe of the flavour content of the proton is provided by the
weak neutral axial current, for which the operator is
\begin{eqnarray}
J^{A,Z}_{\lambda} = \overline{\Psi}_{u} \gamma_{\lambda} \gamma_{5}
\Psi_{u} - \overline{\Psi}_{d} \gamma_{\lambda} \gamma_{5} \Psi_{d} -
\overline{\Psi}_{s} \gamma_{\lambda} \gamma_{5} \Psi_{s} \label{jz}
\end{eqnarray}
For elastic low $Q^{2}$ neutral current processes, the
weak neutral axial quark current for definite flavour gives the
coupling constant for the corresponding axial current of the proton
\begin{eqnarray}
< p \vert \overline{\Psi}_{q} \gamma_{\lambda}
\gamma_{5} \Psi_{q} \vert p > \tendeix{ G_A^q \;\overline{p}
\gamma_{\lambda} \gamma_{5} p}{Q^2}{0}
\end{eqnarray}
Therefore, from the combination Eq.(\ref{jz}) for quark neutral weak
axial current, the proton coupling can be inferred:
\begin{eqnarray}
G_{A} & = & \Delta u - \Delta d - \Delta s = 1.44 \pm 0.06\nonumber
\end{eqnarray}
when the already mentioned analysis is used.  In this
way we are assuming that, although in a different theoretical frame,
we have an {\it a priori} estimate for the coupling constant $G_A$ of
the proton and for the flavour couplings $G_A^q$.  We will use the
values inferred in this way as a guess to compute the observables we
construct in the next section.

In terms of nucleonic isospin, these experiments predict that, in
 addition to the well known isovector axial coupling
 \begin{eqnarray}
 g_{A} & = & \Delta u - \Delta d = 1.254 \pm 0.006 \end{eqnarray}
 an
 isoscalar axial coupling $f_{A}$, such that
 \begin{eqnarray} f_{A} &
 = & - \Delta s = -0.19 \pm 0.06
 \end{eqnarray}
 does exist.

 In this paper we present an extended analysis of the proposal we have
made in Ref.\cite{nos}, for which the neutral vector coupling of the
proton is filtered and only $G_A$ is left in the observables.  The
Primakoff Effect \cite{prim} corresponds to the coherent
photoproduction of $\pi^{0}$ by the nuclear Coulomb field.  This
process is mediated by the axial anomaly \cite{anom} for the
vector-vector-axial current, and the $\pi^0$ field is implemented
using the PCAC hypothesis.  The parity violating asymmetries in the
Primakoff Effect for polarized photons or polarized protons are the
appropriate observables.  In precise terms, the two parity violating
asymmetries for circularly polarized photons or longitudinally
polarized protons are proportional to $G_{A}$, as a result of the
interference of the weak axial neutral current amplitude with the
electromagnetic one, through the magnetic form factor $G_{M}$ or the
electric form factor $G_{E}$ of the proton, respectively.  For more
details we refer to Ref.\cite{nos} and references quoted there.  This
type of analysis is not possible for the $\eta^\prime$.  As it is not
a Goldstone boson there is no legitimate PCAC hypothesis for the
$\eta^\prime$.  A different approach is needed.

If the $u$, $d$ and $s$ quarks were all massless, the low energy
hadron spectrum would consist of a massless $U(3)$ octet of Goldstone
bosons plus a massive singlet, due to the $U(1)$ axial anomaly.  With
non vanishing quark masses the octet becomes massive.  $U(3)$
breaking also causes the singlet $\eta_0$ to mix with the $\eta_8$
producing the physical eigenstates $\eta$ and $\eta^\prime$ to be
\bea
 \ket{\eta} \,=\, \cos\theta\;\ket{\eta_8} - \sin\theta
\ket{\eta_0} &,& \ket{\eta^\prime} \,=\,
\sin\theta\;\ket{\eta_8} + \cos\theta \ket{\eta_0}
\label{mix}
\eea
where no mixing with other $I=0$ pseudoscalar states is
present in the isospin limit.
Equation (\ref{mix}) shows that one can expect the
suppression factor, coming from the anomaly cancellation already
discussed, not to be dominant when considering observables related to
$\eta$ and $\etap$ photoproduction.  This factor certainly appears in
the $\eta_8$ photoproduction, but not in the $\eta_0$ one.  As the
physical $\eta$ has a component along the $\eta_0$
and if the mixing angle is
not too small, this component can prevent the expected values for the
observables to be suppressed.  We will verify that, in fact, this is
the case.

The decay amplitude of a pseudoscalar meson to two photons can be
parametrized as
\beq
 {\cal M}(P_i\lra\gamma (q^\prime)\; \gamma^\ast
(q))=\frac{\alpha}{2\pi\overline{F_i}}\;c_i\;
\epsilon^{\mu\nu\alpha\beta} \;
\varepsilon^\ast_{\mu}\;\varepsilon^{\prime \ast}_\nu\;q_\alpha
q^\prime_{\beta} \,(1 + b_i \,q^2)
\label{ampl}
\eeq
 where
$P_i=(\pi^0,\eta_8,\eta_0)$ and $c_i=(1,\sqrt{\frac{1}{3}},2
\sqrt{\frac{2}{3}})$.  We allow one of the two photons $\gamma^\ast
(q)$ to be off shell; $q^2=0$ when the two photons are on shell.  The
slope parameter $b_i$ gives the leading order term in a
$q^2$-expansion when one photon is off shell.  The axial anomaly plus
PCAC predicts $\overline{F}_\pi=F_\pi$, with a good fit to the $\pi^0$
decay rate.  A powerful theoretical approach
to implement these symmetry features is to refer to the
Wess-Zumino-Witten effective lagrangean \cite{wzw}.  Anomalous
processes are described by this lagrangean.  One loop corrections at
low energies from chiral perturbation theory predict \cite{chpt}
\bea
\overline{F}_{\eta_8}\simeq 1.3 \;F_\pi & \hspace{1.5cm}&
\overline{F}_{\eta_0}\simeq 1.1 \;F_\pi
\eea
while $\theta\simeq
-20^o$.  Assuming that nonet symmetry gives a good description of the
singlet except for its mass which gets an extra term, we have that for
low energy the above formulas are valid.  This argument is supported
\cite{wi} by large $N_c$ arguments that show that, in despite of the
axial anomaly present in the Ward identity for the singlet current, in
that limit the $\eta_0$-analogous is a Goldstone boson.  When one
of the two photons is not on shell, the slope parameter $b_i$ has, in
principle, to be included in the above analysis.  This has been
measured \cite{slope} in the processes $\gamma\gamma^\ast\lra P_i$ in
electron-positron collision and in $\eta \lra \gamma\mu^+\mu^-$.
These measurements are for rather large values of $-q^2$ and a
extrapolation to small values of $q^2$ is needed.  As we are
interested only in low $q^2$ (\ie $\abs{b_i q^2} < 1$), where the
coherent cross section is appreciable, we neglect this term in
Eq.(\ref{ampl}).
The amplitude for the $\gamma - \gamma^\ast -\eta$ and $\gamma -
\gamma^\ast -\eta^\prime$ vertex is then:
\beq
 {\cal M}(\eta \,(\etap
) \lra \gamma\gamma^\ast) = \frac{\alpha}{2\pi F_{\eta \,(\etap )}}\;
\epsilon^{\mu\nu\alpha\beta} \;
\varepsilon^\ast_\mu\;\varepsilon^{\prime\ast}_\nu\;q_\alpha
q^\prime_\beta
\eeq
 where the $\eta$ and $\eta^\prime$ decay constants
are
\bea
 F_{\eta} =
\displaystyle{\left(\frac{1}{\sqrt{3}}\frac{\cos\theta}
{\overline{F}_{\eta_8}} - 2 \sqrt{\frac{2}{3}}
\frac{\sin\theta}{\overline{F}_{\eta_0}}\right)^{-1}}
&,&F_{\eta^\prime} =
\displaystyle{\left(\frac{1}{\sqrt{3}}\frac{\sin\theta}
{\overline{F}_{\eta_8}} + 2 \sqrt{\frac{2}{3}}
\frac{\cos\theta}{\overline{F}_{\eta_0}}\right)^{-1}}
\eea
So we have
a way to compute the polarized Primakoff Effect for the $\eta$ and
$\eta^\prime$.  From now onwards, we proceed along the same method as
in Ref.\cite{nos}.  In the $\eta\;(\eta^\prime )$ Primakoff production
there exists the conventional electromagnetic contribution plus a
neutral weak current contribution through the $\gamma -Z-\eta
\;(\eta^\prime)$ vertex.  In the case of $\pi^0$ the $\gamma - Z$
anomaly is proportional to
\beq
 D^{\gamma Z}_\pi = \frac{N_{c}}{s_{w}
c_{w}} Tr \left[ \left\{ Q^{em}, V^{Z} \right\}
\frac{\lambda_{3}}{2}\right] = \frac{1 - 4 \sin^2\theta_W}{4
\sin\theta_W \cos\theta_W}
\eeq
giving a suppression factor
proportional to the neutral vector coupling $v^e$ of the electron.
Taking into account the mixing given in Eq.(\ref{mix}), the
corresponding coefficients for the $\eta$ and $\eta^\prime$ are
\bea
D^{\gamma Z}_{\eta} = \cos\theta \;D^{\gamma Z}_{\eta_8} -
\sin\theta\; D^{\gamma Z}_{\eta_0}&,& D^{\gamma Z}_{\eta^\prime} =
\sin\theta \;D^{\gamma Z}_{\eta_8} + \cos\theta\; D^{\gamma
Z}_{\eta_0}
\eea
where
\bea
 D^{\gamma Z}_{\eta_0} = \frac{2
\,(1-2\sin^2\theta_W)} {\sqrt{3}\sin\theta_W\cos\theta_W}
&,& D^{\gamma Z}_{\eta_8} =
\frac{1-4\sin^2\theta_W}{4\sqrt{3}\sin\theta_W \cos\theta_W}
\eea
and
the suppression factor for the $\eta_8$ is seen in the last
coefficient.  Finally we get
\beq
D^{\gamma Z}_{\eta} =
\frac{1}{\sqrt{3} \sin\theta_W \cos\theta_W} \left(-\sin^2\theta_W\, (
\cos\theta-4\sin\theta )- 2 \sin\theta + \frac{\cos\theta}{4} \right)
\label{deta}\eeq and
\beq
 D^{\gamma Z}_{\eta^\prime} =
\frac{1}{\sqrt{3} \sin\theta_W \cos\theta_W} \left(-\sin^2\theta_W\, (
\sin\theta+4\cos\theta )+ 2 \cos\theta + \frac{\sin\theta}{4} \right)
\label{detap}
\eeq
\section{Polarization Observables}

One has to look for parity-violating asymmetries in order to
disentangle the $Z$-exchange contribution.  Parity-violating
observables will be induced by the weak-electro\-mag\-ne\-tic
interference for polarized photons or polarized protons.  The
parity-violating interference automatically selects the weak neutral
{\it axial} current of the proton, with coupling $G_{A}$.

For the processes $\gamma (k)\, p \lra \eta \,p^\prime$ and $\gamma
(k)\,p \lra \etap \,p^\prime$ all the observable quantities of
interest are obtained, at lowest order, from the electromagnetic and
weak amplitudes:
\begin{eqnarray}
&&T_\gamma=\frac{ie^3}{2\sqrt{2}\pi^2F_{\eta^\prime} q^2}\;
\epsilon_{\mu\, \nu\, \alpha\, \beta} \;{\cal\epsilon}^\mu(k)
<p^\prime |J_{e.m.}^\nu |p>k^\alpha q^\beta \nonumber\\
&&T_Z=-\frac{ie}{2\pi^2F_{\eta \,(\eta^\prime )}} D^{\gamma Z}_ {\eta
\,(\eta^\prime )}\;G_F\; \sin\theta_W \cos\theta_W\; \epsilon_{\mu\,
\nu\, \alpha\, \beta}\; {\cal\epsilon}^\mu(k) \;<p^\prime
|J_{A,Z}^\nu|p> k^\alpha q^\beta
\label{amplit}
\end{eqnarray}
where
$k$, $p^\prime$ and $p$ are the four-momenta for the incident photon
and the final and initial proton respectively, and $q=p^\prime-p$.
The Lorentz decomposition of the matrix elements is
\begin{eqnarray}
&& <p^\prime |J^\nu_{e.m.}|p>= \overline{u}(p^\prime )\left[
\gamma^\nu F_1(q^2)+i \sigma^{\nu
\mu}\frac{q_\mu}{2M}F_2(q^2)\right]u(p)\nonumber\\ &&<p^\prime
|J^\nu_{A,Z}|p>= G_A(q^2)\;\overline{u}(p^\prime )\gamma^\nu
\gamma_5u(p) + G_P(q^2)\; \overline{u}(p^\prime ) q^\nu \gamma^5 u(p)
\label{axial}
\end{eqnarray}
We notice that the pseudoscalar form
factor $G_P$ of $<p'|J^\nu_{A,Z}|p>$ will be omitted in the following.
This is so because it exactly cancels in the $T_{Z}$ amplitude when
contracted with the anomalous $\gamma -Z-\eta\, (\eta^\prime )$
vertex, as seen in Eq.(\ref{amplit}), in such a way that we do not
have to postulate any extra hypothesis related to this rather unknown
form factor:  it just disappears from the amplitude in these
processes.

For the parity-violating observables we are interested in, the squared
amplitude $|T|^2$ is given, at leading order and at low energies, by
the electromagnetic term plus the electromagnetic-weak interference,
that can be written in the following way:
\begin{equation}
\abs{T}^2 = 8\pi\alpha\
\left(\frac{\alpha}{\pi F_{\eta\,(\eta^\prime )}}\right)^2\, L^{\nu
\mu} \left\{ W^{e.m.}_{\nu \mu} - \sin\theta_W\,
\cos\theta_W\;D^{\gamma Z}_{\eta\,(\etap )}\;
\frac{G_{F}}{\sqrt{2}\,\pi} \frac{q^{2}}{\alpha} W^{I}_{\nu \mu}
\right\}
\label{tensor}
\end{equation}
In order to clarify the
discussion, let us decompose each tensor in Eq.(\ref{tensor}) in two
pieces with definite properties:
\begin{equation}
L^{\nu \mu} =
L^{\nu \mu}_{S} + i L^{\nu \mu}_{A} (h)
\label{tensorl}
\end{equation}
\begin{equation}
W^{e.m.}_{\nu \mu} = W^{e.m.}_{\nu \mu, S} + i
W^{e.m.}_{\nu \mu, A} (s)
\label{tensorw}
\end{equation}
\begin{equation}
W^{I}_{\nu \mu} = i W_{\nu \mu, A}^{I} + W_{\nu
\mu}^{I} (s)
\label{tensorwi}
\end{equation}
Let us summarize the
properties of the different tensors in the above expressions.  The
non-baryonic tensor $L^{\nu \mu}$ is a common factor to
Eq.(\ref{tensor}).  $L^{\nu \mu}_{S}$ is real, symmetric and
independent of the photon helicity $h$, whereas $i L^{\nu \mu}_{A}$ is
imaginary, antisymmetric and linear in $h$.  The two pieces of the
electromagnetic $\gamma - \gamma$ tensor $W^{e.m}_{\nu \mu}$ for the
proton have the following characteristics:  $W^{e.m.}_{\nu \mu,S}$ is
real, symmetric and independent of the proton polarization $s$; and $i
W^{e.m.}_{\nu \mu, A}(s)$, on the contrary, is imaginary,
antisymmetric and linear in $s$.  The interference $\gamma - Z$ tensor
$W^{I}_{\nu \mu}$ for the proton has the following structure:  $i
W_{\nu \mu, A}^{I}$ is imaginary, antisymmetric and independent of the
proton polarization $s$ and, as we will see, in our case it is given
by the axial-magnetic interference; finally $W_{\nu \mu}^{I} (s)$ is
real and linear in $s$.

 For elastic proton scattering and in the laboratory frame, all of
them can be explicitly computed using Eqs.(\ref{amplit},\ref{axial}),
and the result is:
\begin{eqnarray}
&&L^{\nu \mu}_{S}=
\frac{1}{2}\left[ (k^\nu q^\mu+k^\mu q^\nu)(kq)-k^\nu k^\mu q^2-
g^{\nu \mu} (kq)^2\right]\label{lnumus}\\ &&L^{\nu \mu}_{A}
(h)=-\frac{h}{2}\,\epsilon^{\nu \mu\alpha\beta} \;k_\alpha\, q_\beta\;
(kq)
\label{lnumua}
\end{eqnarray}
\beq
W^{e.m.}_{\nu \mu,
S}=G_M^2(q^2g_{\nu \mu}-q_\nu q_\mu)+
(2p_\nu+q_\nu)(2p_\mu+q_\mu)\frac{G_E^2-\frac{q^2}{4M^2}G_M^2}
{1-\frac{q^2}{4M^2}}
\eeq
\beq
W^{e.m.}_{\nu \mu, A}(s)=\epsilon_{\nu
\mu \alpha \beta}q^\alpha \left[ G_M G_E\; (2M)
s^\beta+\frac{(qs)}{M\left(1-\frac{q^2}{4M^2}\right)} \,
G_M(G_M-G_E)\, p^\beta\right]
\eeq
\beq
W_{\nu \mu, A}^{I}= -2\,G_A\,
G_M\; \epsilon_{\nu \mu \alpha \beta}\,q^\alpha p^\beta \eeq
\begin{eqnarray}
W^I_{\nu \mu}(s)&=&2M\,G_A\times \nonumber\\ &&
\left[G_M\left(g_{\nu \mu}(sq)-s_\nu q_\mu\right)
-(2p_\nu+q_\nu)\left(G_Es_\mu+\frac{G_M-G_E}{1-\frac{q^2}{4M^2}}\,
\frac{(sq)}{2M^2}p_\mu\right)\right]
\label{wnumus}
\end{eqnarray}
where $G_E$ and $G_M$ are the two Sachs form factors of the proton:
\bea
G_{E}=F_1+\frac{q^2}{4M^2}F_2 &\hspace{1cm}& G_{M}=F_1+F_2
\eea
We see that the contraction $L^{\nu \mu} W^{e.m.}_{\nu \mu}$ cannot
induce {\it separate} linear terms in $h$ or $s$.  As our aim is the
extraction of $G_{A}$ in $W^{I}_{\nu \mu}$, one can first get the
information from the sector \[ L^{\nu \mu}_{A} (h) \; W_{\nu \mu,
A}^{I} \; , \] by considering, in the laboratory frame, the difference
of cross sections for different photon helicity and for unpolarized
proton
\begin{eqnarray}
\frac{d\sigma(h=+)}{dq^2}-\frac{d\sigma(h=-)}{dq^2}&=&
\frac{\alpha^2}{16\pi^3 F_{\eta\,(\eta^\prime )}^2}\, D^{\gamma Z}_
{\eta\,(\eta^\prime )}\, \frac{G_F \sin\theta_W \cos\theta_W} {\,
ME}\times \nonumber\\ && G_A G_M\left(1-\frac{q^2-m_{\eta\,(\etap
)}^2}{4ME}\right) (q^2-m_{\eta\,(\etap )}^2)
\label{dsg}
\end{eqnarray}
 where $E$ is the photon energy.  The associated
parity-violating observable for circularly polarized photons
corresponds to the following asymmetry:
\beq
A^{\gamma} \equiv
\frac{d \sigma (h=+) - d \sigma (h=-)}{d \sigma (h=+) + d \sigma
(h=-)}
\label{ag}
\eeq
{}From Eqs.(\ref{lnumus},\ref{wnumus}) one can
build a second parity-violating observable from \[ L^{\nu \mu}_{S} \;
W_{\nu \mu}^{I} (s) \; , \] which corresponds to the differences of
cross sections for different proton polarizations and for unpolarized
photons
\begin{eqnarray}
&&\hspace{-1.7cm}\frac{d\sigma(s=+)}{dq^2}-\frac{d\sigma(s=-)}{dq^2}=
\frac{\alpha^2}{8\pi^3 F_{\eta\,(\etap )}^2}\, D^{\gamma Z}_
{\eta\,(\etap )}G_F\sqrt{2} \sin\theta_W \cos\theta_W\,
\frac{1}{1-\frac{q^2}{4M^2}}\, \frac{1}{q^2}\; \times \nonumber\\
&&\hspace{-1.7cm}G_A\left[ G_E\left(1+\frac{q^2-m_{\eta\,(\etap
)}^2}{4ME}\right)\left(q^2+q^2 \frac{q^2-m_{\eta\,(\etap )}^2}{2EM}+
\frac{(q^2-m_{\eta\,(\etap )}^2)^2} {4E^2}\right)-\right.\nonumber\\
&&\hspace{-1.7cm}\left.\frac{G_M}{4}\left(\frac{q^2-m_{\eta\,(\etap
)}^2}{EM} +\frac{q^2} {M^2}\right) \left(q^2+\frac{q^2-m_{\eta\,(\etap
)}^2}{2E}\left(\frac{q^2}{M}+q^2 \frac{q^2-m_{\eta\,(\etap
)}^2}{4M^2E} -\frac{q^2-m_{\eta\,(\etap )}^2}{2E}\right)\right)\right]
\label{dsp}
\end{eqnarray}

The corresponding asymmetry for longitudinally polarized protons is
then given by
\beq
A^{p} \equiv \frac{d \sigma (s=+) - d\sigma
(s=-)}{d \sigma (s=+) + d \sigma (s =-)}
\label{ap}
\eeq
In the next
section the numerical estimate for these two asymmetries $A^\gamma$
and $A^p$ is given.  The above formulas Eq.(\ref{dsg}) and
Eq.(\ref{dsp}) show that our aim is achieved:  both asymmetries are
proportional to the coupling $G_A$. Besides,  the suppression factor $v^e$
is not the leading term in $D^{\gamma Z}_{\eta\,(\etap )}$, as it was
in  $D^{\gamma Z}_{\pi}$.

\section{Numerical Estimates and Conclusions}

For the value of $G_A$ suggested by the EMC-experiment we show in
Figure 1 the expected results of the two asymmetries $A^\gamma$ and
$A^p$ as functions of $-q^2$ from $0.1$ to $0.5$ GeV$^2$, for various
incident energies, and for $\eta$ and $\etap$ photoproduction.
\vspace*{-2cm}\mxfigura{9cm}{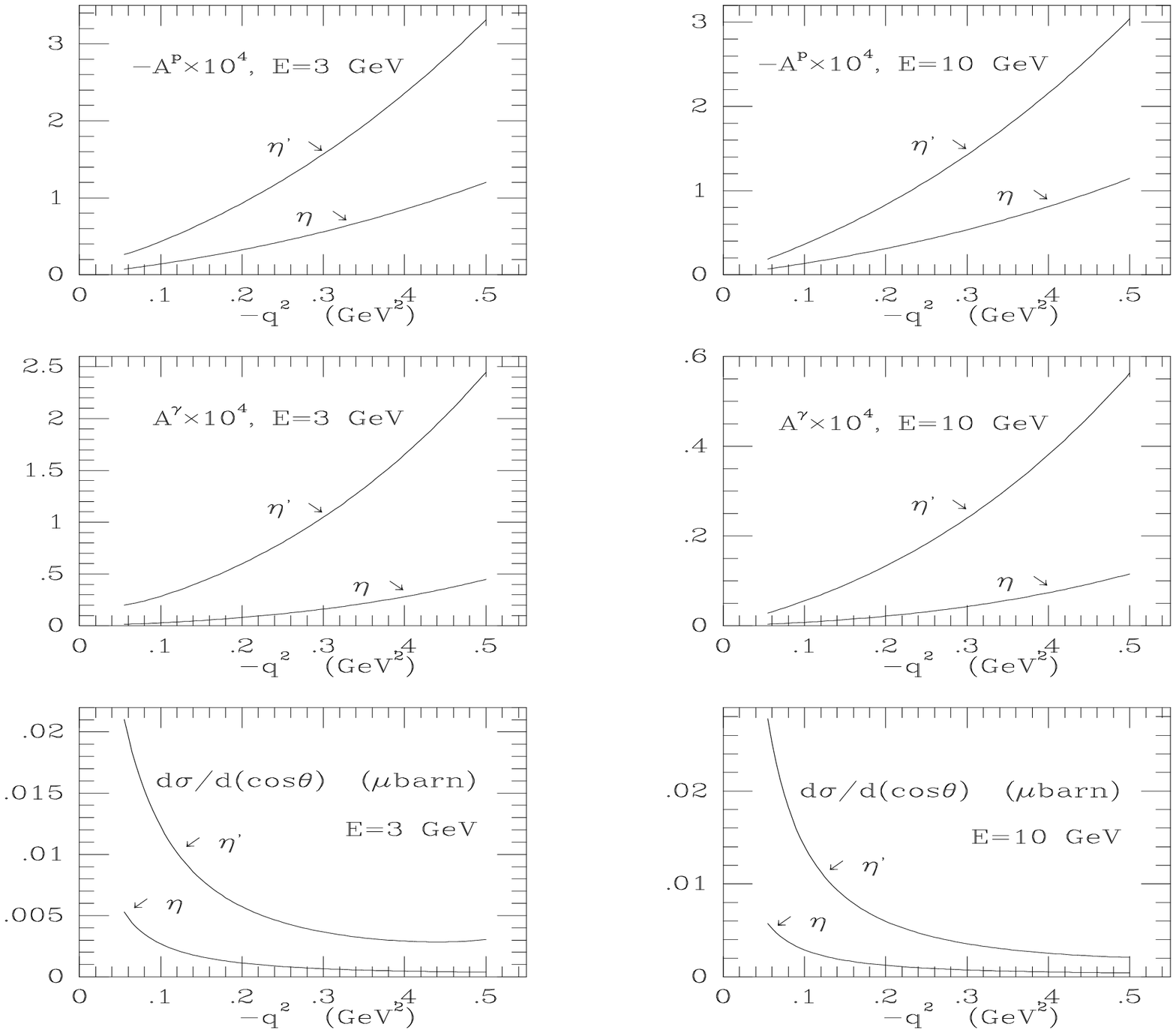}{$\eta$ and $\etap$ polarized
Primakoff Effect.
Cross sections and asymmetries for polarized photon and proton for
photon energies 3 GeV and 10 GeV.}{Figure 1}
 We
also show the cross sections for the considered processes.  The
predictions for the asymmetries are two orders of magnitude bigger
than the ones we have previously obtained in \cite{nos} for the
$\pi^0$.  The minimum
number of events needed to be sensitive to these asymmetries is thus
reduced by two or three orders of magnitude.
 In the case of the $\eta$ this enhancement is due to the
combined effect of a large enough mixing angle that allows the
$\eta_0$ component in Eq.(\ref{mix}) to play a leading role in the
observables so that
an amplification is produced in the anomaly factor:
$D^{\gamma Z}_{\eta}\simeq 12.4 \,D^{\gamma Z}_{\pi}$.  For $\etap$
the enhancement is mostly due to the fact that the $\eta_0$ is the
leading component in Eq.(\ref{mix}) and the anomaly factor is
$D^{\gamma Z}_{\etap}\simeq 32.3 \,D^{\gamma Z}_{\pi}$.

We calculate parity violating asymmetries for polarized photon
Eq.(\ref{ag}) or polarized proton Eq.(\ref{ap})
in the $\eta$ and $\etap$
Primakoff Effect.  They filter the couplings of the proton so as to
leave the weak neutral {\it axial} coupling $G_{A}$ in the
observables, and the contribution coming from the pseudoscalar form
factor $G_P$ is exactly zero.  These asymmetries, due to the
interference between $\gamma$- and Z-exchanges, are mediated by the
$\gamma - Z - \eta \,(\eta^\prime )$ anomaly.  Thus the suppression
factor due to the anomaly cancellation condition, that appears in the
asymmetries for $\pi^0$ and $\eta_8$ photoproduction, is avoided.  The
$\eta$ and $\eta^\prime$ are implemented as a mixing of the
$\ket{\eta_8}$ and $\ket{\eta_0}$ $U(3)$ states, whereas the vertex
is calculated in chiral perturbation theory.  When the value of $G_A$
as determined by the EMC-experiment is used, the predictions for the
asymmetries are of order $10^{-4}$.

\section*{Acknowledgements}

G.A.G.S.  thanks the Spanish Ministerio de Educaci\'on y Ciencia
and the Generalitat Valenciana for a
postdoctoral grant at the University of Valencia.
We acknowledge discussions with J.Bijnens and
A.Pich on the topics of this paper.  This work has been supported in
part by CICYT, under Grant AEN 93-0234, and by I.V.E.I..

\end{document}